# An LLM-Powered Clinical Calculator Chatbot Backed by Verifiable Clinical Calculators and their Metadata

Niranjan Kumar, BS[1], Farid Seifi, PhD[1], Marisa Conte, MLIS[1], Allen J. Flynn, PharmD, PhD[1]
[1]University of Michigan Medical School, Ann Arbor, MI

**Abstract**

*Clinical calculators are widely used, and large language models (LLMs) make it possible to engage them using natural language. We demonstrate a purpose-built chatbot that leverages software implementations of verifiable clinical calculators via LLM tools and metadata about these calculators via retrieval augmented generation (RAG). We compare the chatbot's response accuracy to an unassisted off-the-shelf LLM on four natural language conversation workloads. Our chatbot achieves 100% accuracy on queries interrogating calculator metadata content and shows a significant increase in clinical calculation accuracy vs. the off-the-shelf LLM when prompted with complete sentences (86.4% vs. 61.8%) or with medical shorthand (79.2% vs. 62.0%). It eliminates calculation errors when prompted with complete sentences (0% vs. 16.8%) and greatly reduces them when prompted with medical shorthand (2.4% vs. 18%). While our chatbot is not ready for clinical use, these results show progress in minimizing incorrect calculation results.*

**Introduction**

Large language models (LLMs) have emerged as a powerful generative AI tool for question-answering and summarization with novel text generation.[1] After its public debut in November 2022, the web application ChatGPT[2] not only helped popularize LLMs but also demonstrated their ability to support a responsive, easy-to-use natural language user interface called a *chatbot*. Through chat conversations, LLM-powered chatbots provide access to information and fulfill requests to complete new computations. Researchers have explored the use of LLM-powered chatbots to provide clinical decision support.[3-5] Clinical calculators are a widely used[6] form of clinical decision support valued by providers.[7-9] While some clinical calculators are integrated into electronic health records, clinicians must often navigate to external web-based services (e.g., MDCalc, ClinCalc.com), find an appropriate calculator and manually enter parameter values to complete a calculation. We believe that LLM-powered chatbots have the potential to improve clinician efficiency by providing a natural language user interface for discovering, accessing, and using clinical calculators.

However, reliably performing mathematical calculations remains a challenge for LLMs,[10,11] and this must be overcome before LLM-powered chatbots can be trusted as clinical calculators in patient care. Goodell et al. have shown that OpenAI's ChatGPT version GPT-4 is an unreliable clinical calculator,[12] and the opacity of commercially available LLM-powered chatbots obscures the reasons for their lack of reliability. Typically, users of LLM-powered chatbots like OpenAI's ChatGPT, Microsoft's Copilot, or Google's Gemini have no way of knowing how these systems fulfill clinical calculation requests. Was the correct equation used to compute a patient-specific calculation in a chatbot answer? Were mandatory and optional parameters differentiated? Were units of measure for each parameter made clear and handled properly? Would the LLM chatbot provide the same numeric calculation result regardless of how the query was phrased? To satisfactorily address these kinds of questions, we believe LLM-powered chatbots for clinical use must access and use identifiable, proven software implementations of clinical calculators whose code can easily be isolated, inspected, and tested. We call such proven software implementations ***verifiable clinical calculators***.

We report on the design and testing of a purpose-built LLM-powered chatbot that leverages retrieval augmented generation (RAG)[13,14] and LLM tools[12,15] to provide a natural language interface to find, access, and use *verifiable clinical calculators*. We go beyond adding verifiable clinical calculators to a chatbot by having our chatbot leverage clinical calculator ***metadata*** to answer relevant questions which may be of interest to system administrators, informaticians, and clinicians about each of its verifiable clinical calculators. As a demonstration, we draw comparisons between our new chatbot, backed by verifiable clinical calculators and their metadata, versus a plain chatbot using only *gpt-4o*, an off-the-shelf LLM from OpenAI. Finally, we outline additional future work needed to make clinical calculator chatbots viable in real-world clinical practice.

**Background and Significance**

During their training, clinicians learn how and when to use a variety of biomedical equations while providing patient care. Clinical calculators are commonly used by a wide variety of clinicians,[16] and for a wide variety of purposes—

MDCalc provides access to more than 200 different clinical calculators.[17] As laboratory measurement improves, and as clinical data availability expands, we anticipate that the number and utilization of clinical calculators will grow.

*Calculator Metadata*
Clinical calculators are generally backed by published scientific evidence and usually come with instructions for safe and appropriate use. Key documentation includes the parameters (i.e., inputs, factors, features, variables) found in a clinical equation and their units of measure. Taken together, this content can be thought of as **clinical calculator metadata** describing the calculator's purpose, evidence, use, and parameters. We organize clinical calculator metadata in a standardized linked-data format following recommended practices.[18] Our chatbot accesses this metadata to both perform calculations and respond to queries about the calculators it makes available. The metadata is comprehensive and serves multiple audiences, including clinicians, health IT system administrators, and informaticians.

*RAG and LLM-Extending Tools Mitigate LLM Knowledge Limitations and Hallucinations*
Two main factors limit an LLM's ability to accurately answer questions. First, LLMs are unaware of source material produced after their training cutoff dates, and the training costs makes it impractical to update LLMs frequently enough to keep pace with the publication of new biomedical knowledge. Second, LLMs are prone to *hallucinations*, in which they unintentionally yet confidently produce false statements.[19,20] Even a low rate of hallucinations makes an LLM unfit for clinical use. Retrieval Augmented Generation (RAG)—where a database of curated content retrievable by similarity search with a given query supplements an LLM's intrinsic knowledge[13,14]—can mitigate both problems, and improve LLM performance for clinical decision support.[21] We use RAG to process user queries and clinical calculator metadata associated with 11 verifiable clinical calculators incorporated into our chatbot (Figure 1). Even when RAG is used to supplement LLM knowledge, accurately performing mathematical calculations remains a relative challenge for unassisted LLMs.[15,22,23] The two main approaches to overcoming this challenge are chain-of-thought reasoning[24] and LLM-extending tools.[12,15] We prefer LLM-extending tools because they allow an LLM-powered chatbot to delegate computation to verifiable clinical calculators, potentially eliminating the most concerning LLM hallucinations: those that provide an incorrect calculation result. Goodell et al. have shown that LLM-extending tools can improve the accuracy of clinical calculations with LLM-powered chatbots.[12]

*Verifiable Clinical Calculators*
In addition to eliminating mathematical hallucinations, providing verifiable implementations of clinical calculators as tools to an LLM-powered chatbot imbues calculations with greater trustworthiness. Our approach is informed by ten years of work in the Knowledge Systems Laboratory at the University of Michigan Medical School on metadata for packaging and describing software-based representations of biomedical knowledge, including clinical calculators.[25] Our overall approach to packaging and describing clinical calculators using formally defined and structured *Knowledge Objects*[26] allows our LLM-powered chatbot to guarantee that its calculations are reliably produced using only verifiable clinical calculators in ways that are traceable and transparent. In addition, our approach avoids exposing the code for the software implementations of clinical calculators to any LLM provider (e.g. OpenAI, Microsoft, Google), which is important if clinical calculator logic is proprietary intellectual property. Our approach also facilitates continuous evaluation and updates of clinical calculators and other biomedical algorithms in clinical use.

**Methods**
This study investigates the following research questions: 1) Can an LLM-powered chatbot with access to verifiable clinical calculators and their metadata help clinicians discover and perform calculations with sufficient accuracy for clinical use? 2) How does the accuracy and cost of such a chatbot compare to an off-the-shelf LLM?

*Approach*
We use the following methods to compare the accuracy and performance of our metadata, RAG, and tool-assisted LLM-powered clinical calculator chatbot with an unassisted off-the-shelf LLM-powered chatbot on four simulated natural language conversation workloads. All study materials, including clinical calculator implementations, prompts, chatbot codebase and simulated conversations are available at the project's GitHub site: https://github.com/kgrid-lab/LLM_with_RAG_chatbot.

*Clinical Calculator Implementation*
We incorporated verifiable software implementations of the ten most popular calculators on MDCalc into our chatbot. We chose MDCalc because of its popularity[27] and peer-review process for calculators.[28] MDCalc's list of most popular calculators changes based on usage patterns; in December 2024, it included Creatinine Clearance (Cockcroft-Gault

Equation), CKD-EPI Equations for Glomerular Filtration Rate (GFR), CHA2DS2-VASc Score for Atrial Fibrillation Stroke Risk, Mean Arterial Pressure (MAP), ASCVD (Atherosclerotic Cardiovascular Disease) 2013 Risk Calculator from AHA/ACC, BMI Calculator (Body Mass Index and BSA), Calcium Correction for Hypoalbuminemia and Hyperalbuminemia, Wells' Criteria for Pulmonary Embolism, MDRD GFR Equation, and NIH Stroke Scale/Score (NIHSS). These ten most popular calculators per MDCalc yielded eleven verifiable clinical calculators since Body Mass Index and Body Surface Area are two distinct calculations. We implemented each one as a function in Python 3 (The Python Software Foundation, https://www.python.org/) according to published mathematical formulas.[29-36] For each clinical calculator, we also produced a linked metadata file serialized in JavaScript Object Notation Linked Data (JSON-LD) and conformant with the Knowledge Object paradigm developed by our lab.[26]

*How our Chatbot Works*
Our chatbot uses the *gpt-4o* LLM from OpenAI because OpenAI's models have consistently been shown to be superior to models from other vendors in clinical applications.[37-40] We access OpenAI's LLMs through the Chat Completions API. With each invocation of this API, our chatbot sends the LLM a fixed prompt, the live conversation history, the query from the chatbot user, metadata relevant to the query retrieved from the RAG database, and a list of eleven calculator tools representing the available verifiable clinical calculators.

The prompt is written in a clear stepwise manner per generally accepted prompt-engineering practices.[24,41] To create our RAG database we use Qdrant. (Qdrant Solutions GmbH, Berlin, Germany, https://qdrant.tech/) Each entry in the RAG database is one calculator metadata JSON-LD file. The vector key for each entry is an embedding computed from a minified JSON-LD string of that metadata file content using OpenAI's *text-embedding-3-small* embedding model with the default dimensionality of 1536. With each user query, we use *text-embedding-3-small* to compute the embedding of the query and use this embedding to fetch the single most relevant calculator metadata.

We provide the list of eleven calculator tools to the LLM in the format required by OpenAI. They are programmatically generated from our calculator metadata. When the LLM responds to queries, it decides if it is necessary to invoke one of these tools, and if so, which tool to invoke and what parameter values to provide it. When the LLM decides it needs to invoke a tool, it returns a request to use a tool along with the tool name and parameter values. Our chatbot then calls the appropriate verifiable clinical calculator implementation with the provided parameter values in our local Python environment to obtain the result. We use OpenAI's Structured Outputs feature to ensure that all, and only, the necessary parameters are present and that their values are of the correct type, thereby addressing the potential security vulnerability of executing code with inputs specified by an LLM.

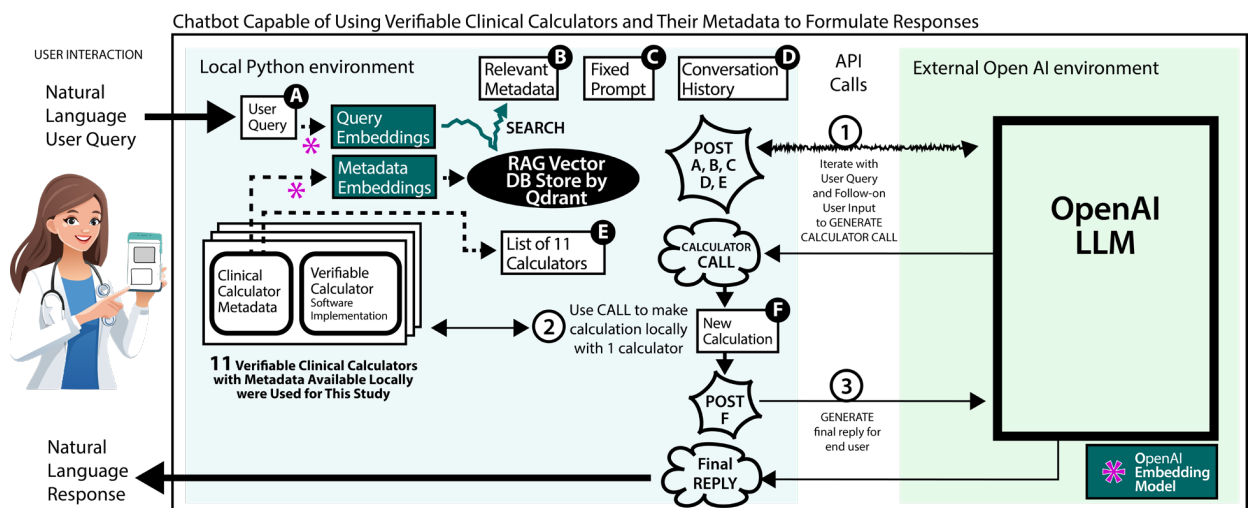

**Figure 1.** Schematic of architecture and data processing patterns of our clinical calculator chatbot.

Putting all these components together (Figure 1), each time our chatbot receives a query, it uses the query to perform a search and retrieve query-relevant calculator metadata (B) and provides these metadata to the LLM along with the current query (A), conversation history (D), and prompt (C). At Step 1, the LLM processes these and determines if the user is requesting to perform a calculation. If not, it responds to the query directly. If the user is requesting a calculation, the LLM identifies which calculator in the provided tool list (E) can perform that calculation and what parameters that calculator requires. If the user does not provide all the required parameters in the initial query, the LLM iteratively prompts the user for any missing parameter values until the user has provided all required parameter values and also provided or explicitly declined to provide all optional parameter values. Then the LLM returns a "Calculator Call", which is a structured request to call the appropriate calculator with the parameter values provided by the user. At Step 2, our chatbot processes the Calculator Call and performs the calculation using a verifiable calculator software implementation. At Step 3, this computed result is returned to the LLM, which generates a final natural language message conveying the result to the user with appropriate context. In this "Final Reply", the LLM clearly delineates verifiable calculator output with asterisks, so it is clear to the user what elements of the chatbot's overall response are from a verifiable calculator and what other elements of the response are generated by the LLM.

*Simulated Conversations*
We created four simulated conversations. The "Full Calculation Conversation" contains queries with fully formed English sentences that ask the chatbot to perform clinical calculations. The "Shorthand Calculation Conversation" is identical in scope to the "Full Calculation Conversation" but uses typical medical shorthand like "pt" for "patient," "yo" for "years old," and "HTN" for "hypertension," in queries that are not complete sentences. Unlike clinical decision support queries presented to LLMs in the literature,[5,12] our queries often deliberately withhold necessary parameter values when initially asking the chatbot to perform a calculation, forcing the chatbot to ask the user for the remaining parameter values, keep track of what information it has received so far, and determine when it has enough information to make a calculation. We believe this is more representative of real-world clinical use as clinicians may not specify every parameter up-front when requesting a calculation in natural language. The other simulated conversations target a system administrator or informatician user. The "Calculator Inquiry Conversation" contains queries which ask the chatbot what each calculator is used for, what the outputs mean, and which input parameters are required or optional, or for similarities and differences between calculators. The "Metadata Inquiry Conversation" contains queries which ask the chatbot technical questions (e.g., what software version of a given calculator is used) that can be answered by referring to clinical calculator metadata.

*Accuracy Evaluation*
We use seven metrics to evaluate the responses of chatbots to our simulated conversations. First, our self-developed "Keyword" metric checks if certain character strings (keywords) are present in the chatbot's response, as most LLM hallucinations involve the omission or mutation of expected output. The second evaluation metric is the Bilingual Evaluation Understudy (BLEU) score.[42] Each query in our simulated conversations has a manually drafted expected response. The BLEU score is a similarity metric computed between this expected response and the chatbot's actual response. Our third, fourth, and fifth metrics are the Recall-Oriented Understudy for Gisting Evaluation with 1-grams (ROUGE-1),[43] with 2-grams (ROUGE-2),[43] and with longest common subsequence (ROUGE-L).[44] Like BLEU, they are computed between the expected and actual response. We use these metrics because they are common in the natural language processing literature.[45,46] Our sixth metric is an LLM-powered response checker which grades each response as either "CORRECT" or "INCORRECT" using the expected response as a reference. Our seventh metric is manual grading of each response, because none of the aforementioned metrics can yet fully substitute for manual grading.

*Performance Evaluation*
For each of the four simulated conversations we gathered performance data including time, cost, and token usage, both on a total and per-query basis. We use Python's datetime module to calculate the compute time durations. We use the OpenAI Usage API to obtain token usage. Since the OpenAI Cost API does not update in real-time, we compute the cost from the token usage according to OpenAI's published rates.

*Quantifying Nondeterminism Arising from LLM Use*
Even with a temperature setting of zero, the output of OpenAI's LLMs remains nondeterministic.[47] In our chatbot's final responses to users, the observed variation is small. We ran each of our conversations through our chatbot five times to quantify this variation. The tables below include the mean and standard deviation of each reported value.

**Results**
*Accuracy*
Tables 1 and 2 present the comparative accuracy results. Our chatbot is significantly more accurate than the unassisted LLM-powered chatbot on all simulated conversation workloads (Full Calculation Conversation: 86.4% vs. 61.6%, P=$5.54\times10^{-7}$; Shorthand Calculation Conversation: 79.2% vs. 62.0%, p=$1.91\times10^{-4}$; Calculator Inquiry Conversation:

100% vs. 75.2%, p undefined; Metadata Inquiry Conversation: 100% vs. 21.0%, p undefined). Notably, our chatbot achieves 100% accuracy across all five trials on the Calculator Inquiry and Metadata Inquiry conversations. The Keyword and LLM evaluation metrics yield similar results to manual evaluation. The BLEU and ROUGE metrics do not, but their results are positively correlated with manual evaluation.

**Table 1. Accuracy of Responses in Conversations Requesting Medical Calculations.** Chatbot accuracy compared to unassisted LLM in simulated conversations requesting calculations (across 5 trials)

| Conversation | "Full Calculation Conversation" | | "Shorthand Calculation Conversation" | |
|---|---|---|---|---|
| | Unassisted LLM | Our Chatbot | Unassisted LLM | Our Chatbot |
| Keyword | 0.636 +/- 0.0261 | 0.820 +/- 0.0141 | 0.600 +/- 0.0245 | 0.772 +/- 0.0672 |
| BLEU | 0.0259 +/- 0.00465 | 0.223 +/- 0.00466 | 0.00999 +/- 0.00184 | 0.147 +/- 0.0163 |
| ROUGE-1 | 0.166 +/- 0.0122 | 0.543 +/- 0.00719 | 0.159 +/- 0.00580 | 0.487 +/- 0.0296 |
| ROUGE-2 | 0.0992 +/- 0.00866 | 0.385 +/- 0.00680 | 0.0709 +/- 0.00253 | 0.291 +/- 0.0184 |
| ROUGE-L | 0.145 +/- 0.0100 | 0.487 +/- 0.00603 | 0.130 +/- 0.00465 | 0.422 +/- 0.0296 |
| LLM | 0.768 +/- 0.0716 | 0.892 +/- 0.0179 | 0.756 +/- 0.00894 | 0.876 +/- 0.0573 |
| Manual | 0.616 +/- 0.0377 | 0.864 +/- 0.00894 | 0.620 +/- 0.0316 | 0.792 +/- 0.0502 |

**Table 2. Accuracy of Responses in Conversations Requesting Calculator Information.** Chatbot accuracy compared to unassisted LLM in simulated conversations requesting information about calculators (across 5 trials)

| Conversation | "Calculator Inquiry Conversation" | | "Metadata Inquiry Conversation" | |
|---|---|---|---|---|
| | Unassisted LLM | Our Chatbot | Unassisted LLM | Our Chatbot |
| Keyword | 0.712 +/- 0.0179 | 1.0 +/- 0.0 | 0.210 +/- 0.113 | 1.0 +/- 0.0 |
| BLEU | 0.0129 +/- 0.00118 | 0.0444 +/- 0.000261 | 0.00420 +/- 0.00247 | 0.0764 +/- 0.0 |
| ROUGE-1 | 0.220 +/- 0.00546 | 0.331 +/- 0.0183 | 0.118 +/- 0.0279 | 0.298 +/- 0.0 |
| ROUGE-2 | 0.111 +/- 0.00327 | 0.203 +/- 0.0134 | 0.0448 +/- 0.0169 | 0.203 +/- 0.0 |
| ROUGE-L | 0.201 +/- 0.00460 | 0.311 +/- 0.0171 | 0.110 +/- 0.0289 | 0.298 +/- 0.0 |
| LLM | 0.832 +/- 0.0335 | 1.0 +/- 0.0 | 0.408 +/- 0.123 | 1.0 +/- 0.0 |
| Manual | 0.752 +/- 0.0179 | 1.0 +/- 0.0 | 0.210 +/- 0.113 | 1.0 +/- 0.0 |

*Failure Mode Analysis*
Table 3 contains the five most common failure modes for our chatbot and the unassisted LLM for the Full Calculation Conversation only. Notably, our chatbot never produces an incorrect calculation result. Its most common errors are neglecting to ask the user to provide values for optional parameters and asking for optional parameter values without saying they are optional. However, on the Shorthand Calculation Conversation, our chatbot occasionally produces incorrect calculation results (Table 4). The incorrect results are all in response to requests to calculate the Wells' Criteria for Pulmonary Embolism. Since our chatbot's LLM delegates computation to a verifiable clinical calculator, there is no error in mathematical operations; rather, the LLM makes incorrect assignments to parameter values.

**Table 3. Top Five Failure Modes in "Full Calculation Conversation."** Top 5 failure modes of our chatbot compared to an unassisted LLM in a simulated conversation with a clinician user requesting to perform calculations.

| Unassisted LLM | | Our Chatbot | |
|---|---|---|---|
| *Failure Mode* | *% Queries Affected* | *Failure Mode* | *% Queries Affected* |
| Incorrect calculation result | 16.8 +/- 3.03 | Fails to ask user if they would like to provide value for optional parameter | 4.00 +/- 0.0 |
| Fails to ask user if they would like to provide value for optional parameter | 5.20 +/- 1.10 | Asks for optional parameter as if mandatory | 3.60 +/- 0.894 |
| Asks for optional parameter as if mandatory | 3.60 +/- 2.19 | Stops prompting user to provide remaining parameters after they provide some | 2.00 +/- 0.0 |
| Fails to ask user for required parameter and assumes value | 2.40 +/- 0.894 | Fails to ask user for required parameter and assumes value | 2.00 +/- 0.0 |
| Fails to clarify which calculator to use among multiple options | 1.60 +/- 0.894 | Fails to clarify which calculator to use among multiple options | 2.00 +/- 0.0 |

**Table 4. Top Five Failure Modes in "Shorthand Calculation Conversation."** Top 5 failure modes of our system compared to unassisted LLM in a simulated conversation in medical shorthand requesting calculations.

| Unassisted LLM | | Our Chatbot | |
|---|---|---|---|
| *Failure Mode* | *% Queries Affected* | *Failure Mode* | *% Queries Affected* |
| Incorrect calculation result | 18.0 +/- 3.74 | Fails to ask user if they would like to provide value for optional parameter | 6.80 +/- 2.68 |
| Fails to ask user if they would like to provide value for optional parameter | 8.40 +/- 0.894 | Does not mention name of calculator | 4.00 +/- 0.0 |
| Refuses to perform calculation | 2.00 +/- 2.83 | Incorrect calculation result | 2.40 +/- 1.67 |
| Fails to ask user for required parameter and assumes value | 2.00 +/- 0.0 | Asks for optional parameter as if mandatory | 2.00 +/- 0.0 |
| Fails to clarify which calculator to use among multiple options | 2.00 +/- 0.0 | Fails to clarify which calculator to use among multiple options | 2.00 +/- 0.0 |

*Performance*

Tables 5 and 6 present performance information, including time, token usage, and cost, both on an overall and per query basis, for both our chatbot and the unassisted LLM. For conversations requesting calculations, our chatbot is faster than the unassisted LLM (Full Calculation Conversation: 4.8 vs. 7.7 seconds per query, Shorthand Calculation Conversation: 4.8 vs. 6.0 seconds per query). For conversations requesting information about calculators, our chatbot is slower than the unassisted LLM (Calculation Inquiry Conversation: 4.0 vs 2.6 seconds per query, Metadata Inquiry Conversation: 2.6 vs 1.3 seconds per query). Our chatbot uses more input tokens per query, both cached and not cached, than the unassisted LLM. Our chatbot uses fewer output tokens per query than the unassisted LLM for the conversations where calculations are requested and roughly the same number for the conversations with queries about the calculators or their metadata. Since both chatbots use far more input tokens than output tokens, our chatbot has an overall higher cost per query than the unassisted LLM by a factor of 2 to 6 times.

**Table 5. Computational Performance and Cost of Conversations Requesting Calculations.** Chatbot performance compared to an unassisted LLM in simulated conversations where clinician users request to perform calculations.

| Conversation | "Full Calculation Conversation" | | "Shorthand Calculation Conversation" | |
|---|---|---|---|---|
| | **Unassisted LLM** | **Our Chatbot** | **Unassisted LLM** | **Our Chatbot** |
| Time (min:sec) | 6:24.511+/-1:47.350 | 3:58.059+/-0:08.530 | 4:59.591 +/- 0:32.977 | 4:01.288 +/- 0:07.520 |
| Time per query (sec) | 7.690 +/- 2.147 | 4.761 +/- 0.171 | 5.992 +/- 0.660 | 4.826 +/- 0.150 |
| Input tokens | 424934 +/- 29994 | 1318802 +/- 25958 | 381731 +/- 21662 | 1334572 +/- 12869 |
| Input tokens per query | 8499 +/- 600 | 26376 +/- 519 | 7635 +/- 433 | 26691 +/- 257 |
| Cached input tokens | 397491 +/- 32853 | 1215002 +/- 29833 | 353178 +/- 19610 | 1238579 +/- 16233 |
| Cached input tokens per query | 7950 +/- 657 | 24300 +/- 597 | 7064 +/- 392 | 24772 +/- 325 |
| Output tokens | 18214 +/- 6734 | 4268 +/- 154 | 14096 +/- 1098 | 4119 +/- 95 |
| Output tokens per query | 364 +/- 135 | 85 +/- 3 | 282 +/- 22 | 82 +/- 2 |
| Cost (USD) | $0.748 +/- $0.0520 | $1.82 +/- $0.0377 | $0.654 +/- $0.0427 | $1.83 +/- $0.0348 |
| Cost per query (USD) | $0.0150 +/- 0.00104 | $0.0364+/- 0.000754 | $0.0131 +/- 0.000854 | $0.0366+/- 0.000696 |

**Table 6. Computational Performance and Cost of Requesting Information About Calculators.** Chatbot's performance compared to unassisted LLM in simulated conversations asking for information about calculators.

| Conversation | "Calculator Inquiry Conversation" | | "Metadata Inquiry Conversation" | |
| --- | --- | --- | --- | --- |
| | Unassisted LLM | Our Chatbot | Unassisted LLM | Our Chatbot |
| Time (min:sec) | 1:04.014+/-0:11.342 | 1:40.819+/-0:11.431 | 1:36.242 +/- 0:04.961 | 3:19.627 +/- 0:11.803 |
| Time per query (sec) | 2.561 +/- 0.454 | 4.033 +/- 0.457 | 1.250+/- 0.064 | 2.593 +/- 0.153 |
| Input tokens | 39889 +/- 1142 | 266546 +/- 1405 | 210918 +/- 5941 | 1549061 +/- 70 |
| Input tokens per query | 1596 +/- 46 | 10662 +/- 56 | 2739 +/- 77 | 20118 +/- 1 |
| Cached input tokens | 25088 +/- 3187 | 240666 +/- 7560 | 173261 +/- 4896 | 1505818 +/- 65334 |
| Cached input tokens per query | 1004 +/- 127 | 9627 +/- 302 | 2250 +/- 64 | 19556 +/- 848 |
| Output tokens | 2307 +/- 168 | 2416 +/- 147 | 2229 +/- 188 | 2172 +/- 1 |
| Output tokens per query | 92 +/- 7 | 97 +/- 6 | 29 +/- 2 | 28 +/- 0 |
| Cost (USD) | $0.0914+/- $0.00373 | $0.390 +/- $0.0129 | $0.333 +/- $0.0145 | $2.01 +/- $0.0816 |
| Cost per query (USD) | $0.00366+/- .000149 | $0.0156+/- 0.000517 | $0.00432+/-0.000189 | $0.0261 +/- 0.00106 |

This study addressed two research questions. We asked whether an LLM-powered chatbot configured with verifiable clinical calculators and their metadata help clinicians discover and perform calculations with sufficient accuracy for clinical use. The answer is no; we discuss our chatbot's limitations below. We asked how the accuracy and cost of a chatbot configured with its own verifiable calculators and their metadata compares to an unassisted, off-the-shelf LLM. Our chatbot's calculation accuracy is higher than the unassisted LLM, and it accurately answers questions about verifiable clinical calculators using metadata. Using our chatbot to perform clinical calculations comes with a 2-6x increase in cost per query relative to the unassisted LLM, and a final per-query cost of 1-4 cents.

**Discussion**

*LLM as Clinical Calculator Interface*
Our clinical calculator chatbot was more accurate than an unassisted off-the-shelf LLM in conversations where a clinician user requests calculations be performed, validating our approach of using metadata, RAG, and LLM tools. Our results go beyond prior work[12] by scaling up to 11 calculators, incorporating calculators with optional parameters, handling conversations where not all required parameter values are provided up-front, and addressing non-determinism inherent in LLMs. For clinical adoption, while accuracy must be near-perfect, with no critical errors, the worst errors are those in which an erroneous calculation result is returned. Our chatbot makes no such errors in the conversation requesting calculations using full sentences, but occasionally makes them in the conversation using medical shorthand. Medical shorthand is challenging, leading to lower overall accuracy for both our chatbot and the unassisted LLM. This is expected since medical shorthand differs from typical natural language enough that most untrained people cannot understand it,[48] and is likely far less prevalent in the training data of commercially available LLMs. To be used in patient care, future clinical calculator chatbots must successfully handle medical shorthand.

*Answering Queries about Calculators using their Metadata*
This study makes a significant contribution by showing how a clinical calculator chatbot can use metadata to answer queries about the calculators themselves. We believe this capability may be of interest to system administrators and informaticians as much or more than it is to clinicians. Our chatbot consistently achieves 100% accuracy in our simulated conversations testing such calculator information queries. We believe this ability is essential in a real-world healthcare organization, where institutional and proprietary algorithms may coexist with widely used calculators, and where implementations of both have to be periodically evaluated and kept up to date.

*Resource Utilization*
Our chatbot demonstrates lower query response latency in conversations where a clinician user requests calculations. This is a promising finding and suggests that high-reliability clinical calculator chatbots will not degrade user experience relative to other chatbots. However, our chatbot exhibits somewhat higher query response latency in conversations with queries about calculators and their metadata. This is to be expected given the extra work to retrieve and incorporate relevant calculator metadata through RAG, and we believe this is an acceptable tradeoff. All response latencies are on the order of seconds using consumer-grade hardware, so this difference may not be meaningful during real-world use. While our chatbot costs 2-6x more than the unassisted LLM per query, the per-query cost is only $0.01 - $0.04, which is likely affordable in the context of overall hospital expenditures.[49]

*Future work*
Our chatbot had some difficulty understanding and responding accurately to queries in medical shorthand. Fine-tuning, which adjusts the weights of an LLM by exposing the LLM to examples representative of the desired use-case,[50] may offer a solution. Fine-tuning has been used with clinical documentation,[51] which typically contains medical shorthand.

One important contribution of this work is the use of LLM tools to delegate computation to verifiable implementations of clinical calculators to eliminate LLM hallucinations affecting mathematical computations. However, an LLM's ability to use tools like clinical calculators decreases as their number increases, and OpenAI recommends using no more than 20 tools[52]. As services like MDCalc have over 200 calculators,[17] this limitation must be overcome before a tool-based calculator chatbot like ours can be deployed clinically. Possible avenues to do this include fine-tuning, reliability alignment,[53] and delegation of tasks by one LLM invocation to others, each of which manages only a subset of tools.

Finally, creation of a chatbot that uses multiple calculators to execute complex tasks is a natural next step. For example, to calculate the bleeding risk of a patient on warfarin, the chatbot would identify a bleeding risk score calculator like ATRIA,[54] recognize GFR as a required parameter, calculate GFR using an appropriate calculator, pass the output to the ATRIA calculator, and return the bleeding risk score to the user. Identifying the appropriate calculator when the user does not explicitly request a calculation remains a challenge for today's LLMs,[55] but our chatbot's 100% accuracy in answering queries about calculators suggests our approach using metadata and RAG could offer a solution.

**Conclusion**
We developed a purpose-built clinical calculator chatbot leveraging calculator metadata, RAG, and an LLM to take advantage of verifiable clinical calculators. We examined its accuracy and performance compared to an unassisted off-the-shelf LLM on four natural language conversation workloads. We find that our chatbot is more accurate than an unassisted LLM in all simulated conversations, achieving 100% accuracy responding to queries about the calculators and their metadata, eliminating incorrect calculation results when clinician-users request calculations using full sentences, and greatly reducing them when requests are made in medical shorthand. These improvements in accuracy come without significant increases in latency but do result in increased compute cost, though the final cost per query is only ~$0.01 to $0.04. We identify overall accuracy, medical shorthand, and scalability of LLM tools as barriers that need to be overcome to make clinical calculator chatbots reliable enough to be used in patient care.